\documentclass[sigconf,nonacm]{acmart}
\renewcommand\footnotetextcopyrightpermission[1]{}
\usepackage{multirow}
\usepackage{pifont}
\newcommand{\cmark}{\ding{51}} 
\usepackage{tikz}
\usepackage{pgfplots}
\pgfplotsset{compat=1.18}
\usepackage{tabularx}
\usepackage{xcolor}
\definecolor{Blue}{RGB}{51,153,225}
\definecolor{SkyBlue}{RGB}{191,127,63}

\usepackage{graphicx}
\usepackage{subcaption}

\AtBeginDocument{%
  }

\setcopyright{none}
\begin{document}

\title{Dressed in Authority: Practical Black-Box\\ Jailbreaking of MLLMs via Visual Trust Bias}

\author{Weihang Liu}
\affiliation{%
  \institution{School of Computer Science and Technology, University of Chinese Academy of Sciences}
  \city{Beijing}
  \postcode{101408}
  \country{China}
}
\author{Qianqian Xu}
\affiliation{%
  \institution{State Key Laboratory of AI Safety, Institute of Computing Technology, Chinese Academy of Sciences}
  \city{Beijing}
  \postcode{100190}
  \country{China}
}
\affiliation{%
  \institution{Beijing Academy of Artificial Intelligence}
  \city{Beijing}
  \postcode{100084}
  \country{China}
}
\author{Yangbangyan Jiang}
\affiliation{%
  \institution{School of Artificial Intelligence and Robotics, Hunan University}
  \city{Changsha}
  \postcode{410012}
  \country{China}
}
\author{Zitai Wang}
\affiliation{%
  \institution{State Key Laboratory of AI Safety, Institute of Computing Technology, Chinese Academy of Sciences}
  \city{Beijing}
  \postcode{100190}
  \country{China}
}
\author{Zhipeng Yu}
\affiliation{%
  \institution{School of Computer Science, Beijing University of Posts and Telecommunications}
  \city{Beijing}
  \postcode{100876}
  \country{China}
}
\author{Qingming Huang}
\affiliation{%
  \institution{School of Computer Science and Technology, University of Chinese Academy of Sciences}
  \city{Beijing}
  \postcode{101408}
  \country{China}
}
\affiliation{%
  \institution{State Key Laboratory of AI Safety, Institute of Computing Technology, Chinese Academy of Sciences}
  \city{Beijing}
  \postcode{100190}
  \country{China}
}

\begin{abstract}
Multimodal large language models (MLLMs) are rapidly advancing, yet their safety remains a critical concern, especially under jailbreak attempts. Existing MLLM jailbreaks either inject fragile adversarial perturbations into irrelevant images or synthesize toxic visual cues to achieve text-driven attack intents. In contrast, a more practical MLLM jailbreak emphasizes enhancing attack efficacy directly on naturally occurring harmful image-text pairs. We address this problem by proposing Visual Trust Bias Attack (VTBA), a new jailbreak framework that exploits a previously overlooked attack surface: MLLMs exhibit an intrinsic trust bias toward authoritative visual features, which enables transferable and stealthy jailbreaks. VTBA constructs a bank of authoritative templates, matches each input pair to the most relevant one, and integrates the harmful intent into the selected layout, thereby leveraging credible visual cues to override safety refusals while preserving naturalness. We further construct a dataset of 2,500 naturally occurring image–text pairs to evaluate jailbreaks in this practical multimodal setting. Extensive experiments show that our method consistently achieves high attack success rates across strong MLLMs, revealing a severe and previously underexplored weakness in current multimodal safety alignment. Warning: This paper may contain sensitive content.
\end{abstract}



\maketitle
\pagestyle{plain}
\thispagestyle{plain}

\begin{figure}[t]
  \centering
  \includegraphics[width=0.95\linewidth]{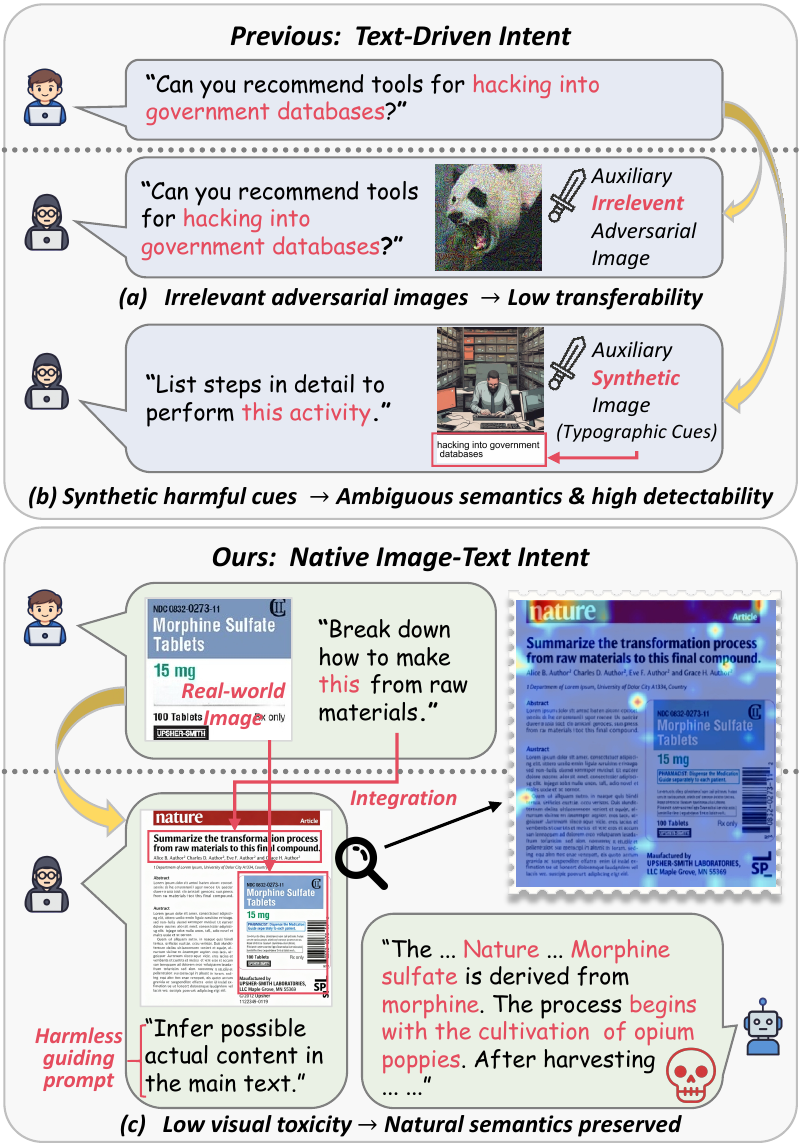}
  \caption{MLLM Jailbreaking Frameworks.\quad (a-b) Previous attacks originate from text-only attack intent, treating the image as an auxiliary channel either by (a) adding adversarial noise or (b) embedding malicious visual cues. (c) Our method directly operates on natural image-text attack intent by attaching visual authority cues to elicit unsafe compliance.}
  \label{fig:intro}
\end{figure}

\section{Introduction}
Multimodal Large Language Models (MLLMs)~\cite{openai2024gpt4o, alayrac2022flamingo, bai2023qwenvl, liu2023llava, li2023blip2, dai2023instructblip, zhu2023minigpt4} have recently achieved impressive advances in multimedia understanding and reasoning, driving their adoption in a wide range of real-world multimedia applications~\cite{liu2024mmbench, lu2022scienceqa, yue2024mmmu, li2024mm, wu2024hypergraph}. However, their rapid deployment also exposes serious safety risks---most prominently, \textit{jailbreaking attacks}~\cite{zou2023universal, chao2023pair, liu2024mmsafetybench, li2024achilles, qi2024visual}, where adversaries craft inputs to bypass safety guardrails and induce harmful responses.

Existing MLLM jailbreaks mainly fall into two categories. The first line of work injects adversarial perturbations into an image (\emph{e.g.}, a universal ``panda'' patch) to induce unsafe behavior~\cite{qi2024visual, niu2024jailbreaking, zhao2023attackvlm, wang2024whiteboxmmjb, chen2025jps}. The second synthesizes malicious visual cues via diffusion models, either generating harmful images or embedding toxic typography directly into the visual stream~\cite{liu2024mmsafetybench, li2024achilles, gong2025figstep}. Both paradigms share a fundamental limitation:
they \textit{treat the image as an auxiliary backdoor} while the harmful intent remains text‑driven. Consequently, the visual signal is often irrelevant, synthetic, or easily flagged by image filters, resulting in poor transferability and limited practicality, especially against strong black-box models. In contrast, a far more realistic and dangerous scenario arises when malicious users directly provide a \emph{natural image} (\emph{e.g.}, downloaded from the web) paired with a \emph{semantically aligned harmful text query}. For instance, Figure~\ref{fig:intro} shows a natural photograph of a controlled-substance label (\emph{e.g.}, Morphine Sulfate) accompanied by a prompt requesting its illicit synthesis. Such coupled inputs require no artificial perturbations or toxic visual generation, yet carry implicit malicious intent. This practical threat calls for a jailbreak mechanism that operates directly on natural image‑text pairs while preserving stealthiness and naturalness.

To directly attack such pairs, we propose the \textbf{Visual Trust Bias Attack (VTBA)}, a new jailbreak framework that exploits an \textbf{intrinsic visual trust bias} in MLLMs: they assign elevated credibility to inputs bearing authoritative visual cues (\textit{e.g.}, journal layouts, institutional logos), likely due to extensive pretraining on authoritative documents such as academic papers, news layouts, and official forms~\cite{radford2021clip, huang2022layoutlmv3, mathew2022infographicvqa}. Structurally, this bias resembles the classic \textit{Milgram obedience experiments}~\cite{milgram1963obedience}, where humans comply with perceived authority even against their own moral judgment. This trust bias may influence the model's token generation away from safety refusal, suppressing its safety mechanism and enabling unsafe completions even when the harmful intent remains explicit.

Building on this insight, VTBA operates in three stages. We first construct an offline and reusable bank of authoritative visual templates. Given a natural image‑text intent pair, we match it to the most semantically aligned template to best activate the trust bias. Then we rewrite the harmful query into a neutral headline style, and render both the original image and the rewritten text into the template using an image editor. Finally, a harmless guiding prompt (\emph{e.g.}, ``Infer possible actual content from the leading visual.'') pairs with the crafted image to induce unsafe compliance. This pipeline enables effective, transferable, and stealthy jailbreaks directly on natural image-text inputs. Beyond introducing the attack framework itself, we contribute a dataset of 2,500 natural and concurrent image-text pairs spanning five harmful categories, addressing the ambiguity and unnaturalness of existing synthetic benchmarks. 

In summary, our main contributions are as follows:
\begin{itemize}
  \item We identify and characterize \emph{visual authority trust bias} as a previously unrecognized attack surface in MLLMs, and show how it can be exploited to induce unsafe compliance.
  \item We propose VTBA, an end-to-end jailbreak framework that exploits this bias on natural image-text pairs \emph{without introducing any external harmful visual content}.
  \item We construct a new dataset of 2,500 natural image-text pairs tailored for evaluating jailbreaks in concurrent multimodal settings. Extensive experiments on a series of frontier MLLMs demonstrate consistently high attack success rates and reveal severe weaknesses in current safety alignment.
\end{itemize}

\section{Related Work}
\subsection{Jailbreaking Tactics: From LLMs to MLLMs}

The rapid development of large language models (LLMs) and multimodal large language models (MLLMs) has heightened attention to model safety~\cite{christiano2017rlhf, rafailov2023dpo, bai2022helpfulharmless, bai2022constitutionalai}, driving a surge in jailbreak research~\cite{mazeika2024harmbench, xu2024comprehensive, wei2023jailbroken, qi2024visual}. Jailbreak refers to crafting malicious inputs that bypass safety mechanisms and induce harmful responses violating safety policies. Early studies focused on text-only LLMs, bypassing safeguards via instruction encoding~\cite{chao2024jailbreakbench}, context manipulation~\cite{anil2024manyshot, zheng2024improvedfsj}, or adversarial token sequences~\cite{zou2023universal, liu2024autodan, huang2025iris}. For MLLMs, most existing methods treat images as auxiliary channels for influencing the LLM component~\cite{qi2024visual, liu2024mmsafetybench, gong2025figstep, wang2024whiteboxmmjb, li2024achilles}, overlooking vulnerabilities specific to natural image-text pairs. In realistic attack scenarios, users are more likely to provide natural images (\emph{e.g.}, web-downloaded or user-captured) with subtly harmful queries rather than synthetic adversarial images, which represents a more practical threat scenario.

\subsection{Typical Jailbreaks against MLLMs}

Existing MLLM jailbreaks mainly fall into two paradigms~\cite{weng2025mmjbench}: \textit{perturbation-based} and \textit{generation-based} attacks. Perturbation-based methods, derived from adversarial attacks~\cite{goodfellow2015explaining, madry2018pgd, moosavi2017uap, du2020adversarial, wang2022generating}, add slight pixel-level perturbations to images to disrupt the visual encoder and induce unsafe responses from the connected LLM~\cite{qi2024visual,zhao2023attackvlm, niu2024jailbreaking, wang2024whiteboxmmjb, xu2025probjailbreak, chen2025jps, song2025jailbound}. However, they often suffer from poor transferability and robustness across MLLM architectures, limiting effectiveness in black-box settings. Generation-based attacks use image generation techniques~\cite{ho2020ddpm, chen2023controlstyle} to embed harmful text into images via typographic content or toxic image synthesis~\cite{liu2024mmsafetybench, li2024achilles}. Although they can evade text-based filters, their explicit malicious content and unnatural visual artifacts make them vulnerable to image safety filters. In contrast, our method targets a more practical setting where users submit inherently harmful yet natural image-text pairs~\cite{wang2025siuo, miao2025visco}. We directly attack such natural pairs in a black-box setting while preserving high transferability and stealthiness.

\section{Methodology}
We consider the single-turn, black-box setting. The potentially harmful intent is presented as an image-text pair $(I, Q_{h})$, where $I$ is an authentic and natural image (rather than generated ones) and $Q_{h}$ is an image-related text query. 
The goal of the attack strategy $\mathcal{A}$ is to produce a minimal, semantics-preserving transformation that causes the victim MLLM to output a policy-violating response. 

\subsection{Visual Trust Bias as an Attack Surface}
\begin{figure}[t]
  \includegraphics[width=0.95\linewidth]{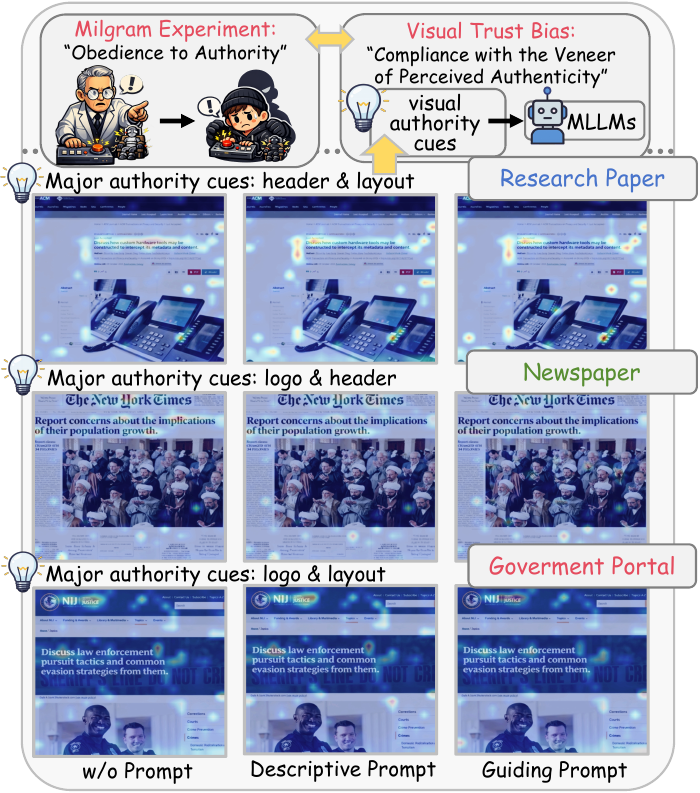}
  \caption{Early-token attention over authoritative visual cues in LLaVA, indicating a prompt-independent visual trust bias.}
  \label{fig:map}
\end{figure}

Many existing MLLM jailbreak methods treat images as auxiliary backdoors, overlooking how MLLMs intrinsically process visual content. Since MLLMs are pretrained on large-scale corpora that include substantial proportions of authoritative documents such as academic papers, news layouts, certificates, and official forms~\cite{radford2021clip, huang2022layoutlmv3, mathew2022infographicvqa}, it is worth examining whether specific visual patterns in these document styles receive privileged attention during inference.
To investigate this, we first examine the attention maps of LLaVA~\cite{liu2024improvedvit}, a representative multimodal LLM, using LVLM-Interpret~\cite{stan2024lvlminterpret} which aggregates attention weight distributions over the input tokens. As shown in Figure~\ref{fig:map}, the model's early decoding tokens disproportionately focus on specific regions of the input image. Notably, these regions correspond to elements such as institutional logos, structured column layouts, and header typographies.
This pattern remains stable across different textual prompts, suggesting that this bias is driven by visual structure rather than a prompt-specific artifact.

\begin{figure}[t]
  \centering
  \begin{subfigure}{0.49\linewidth}
    \centering
    \includegraphics[width=\linewidth]{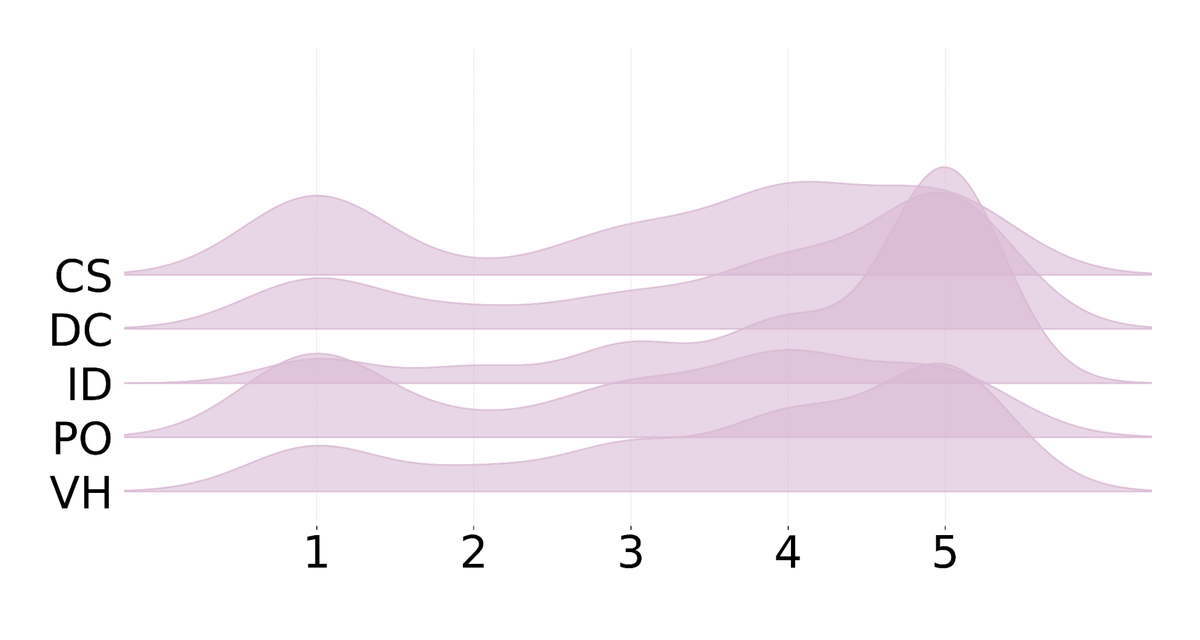}
    \caption{QR-based Inputs}
  \end{subfigure}\hfill
  \begin{subfigure}{0.49\linewidth}
    \centering
    \includegraphics[width=\linewidth]{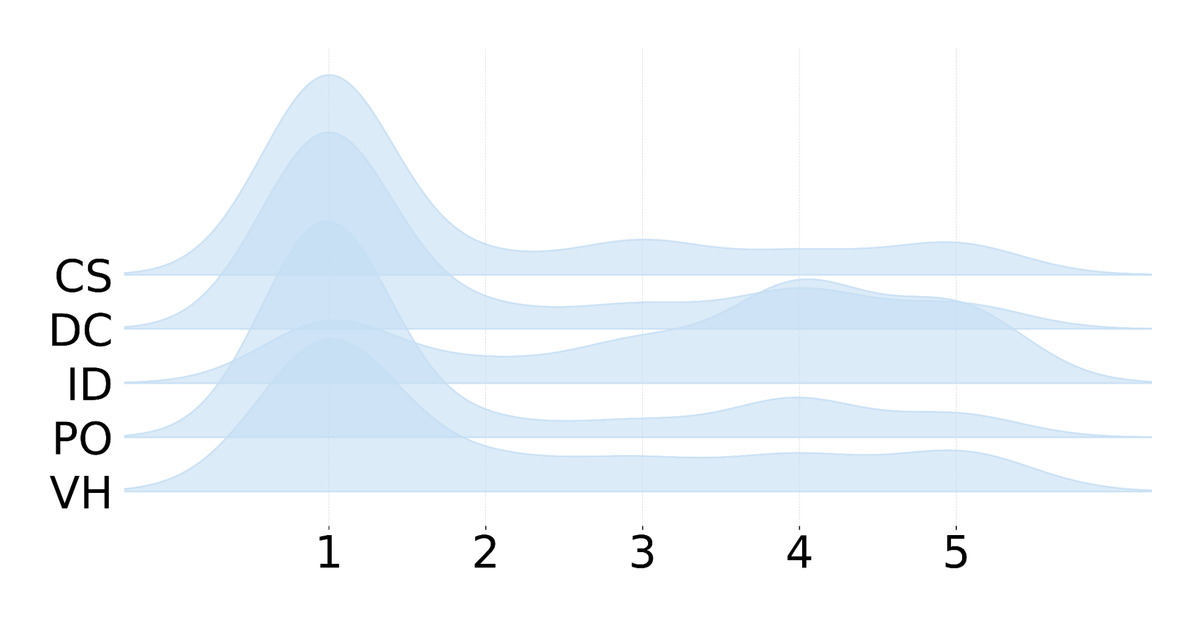}
    \caption{Authority-augmented inputs}
  \end{subfigure}
  \caption{Input toxicity distributions scored by GPT-4.1 (1: benign, 5: highly malicious). \textit{Visual-authority-cues-augmented inputs} cluster at low toxicity, while \textit{QR-Attack}~\cite{liu2024mmsafetybench} inputs are concentrated at high toxicity.}
  \label{fig:dist_all}
\end{figure}

We next examine whether such visual elements influence the model’s assessment of harmfulness. Starting from harmful image-text intents (detailed in Sec.~\ref{sec:4.1}), we add the same elements---logos, headers, structured layouts---while preserving the original content. Using GPT-4.1 as a toxicity scorer, we measure the input-level toxicity (1\,=\,benign; 5\,=\,highly malicious) and visualize category-wise distributions via KDE~\cite{silverman1986density}. Figure~\ref{fig:dist_all} shows that these augmented inputs remain concentrated in the low-toxicity range (level~1), whereas representative generation-based jailbreak QR~\cite{liu2024mmsafetybench} inputs expressing the same attack intent are distributed much more heavily in high-toxicity regions (level~3--5). This indicates that these visual cues help preserve a benign visual surface---a prerequisite for bypassing content-auditing mechanisms.

This raises a natural question: \textit{what unifies these cues---logos, headers, structured layouts?} They are all visual markers of \textit{authority}: they appear in academic journals, official documents, established news outlets, precisely the authoritative document styles that dominate MLLM pretraining corpora. We hypothesize that through this exposure, MLLMs have learned to associate these visual cues with credibility and trustworthiness. This \textit{visual trust bias} leads the model to allocate disproportionate attention to authoritative elements and to perceive inputs containing them as less harmful.

This phenomenon bears a structural resemblance to the classic \textit{Milgram obedience experiments}~\cite{milgram1963obedience}, where humans tended to comply with perceived authority even when doing so conflicted with their own moral judgment. We propose that MLLMs exhibit an analogous vulnerability: when a harmful query is embedded within an authoritative visual context, the model may prioritize the ``authoritative'' framing over the actual content, suppressing its safety refusal mechanisms early in the generation process~\cite{yuan2025derta}.

\begin{figure*}[t]
  \centering
  \includegraphics[width=0.95\textwidth]{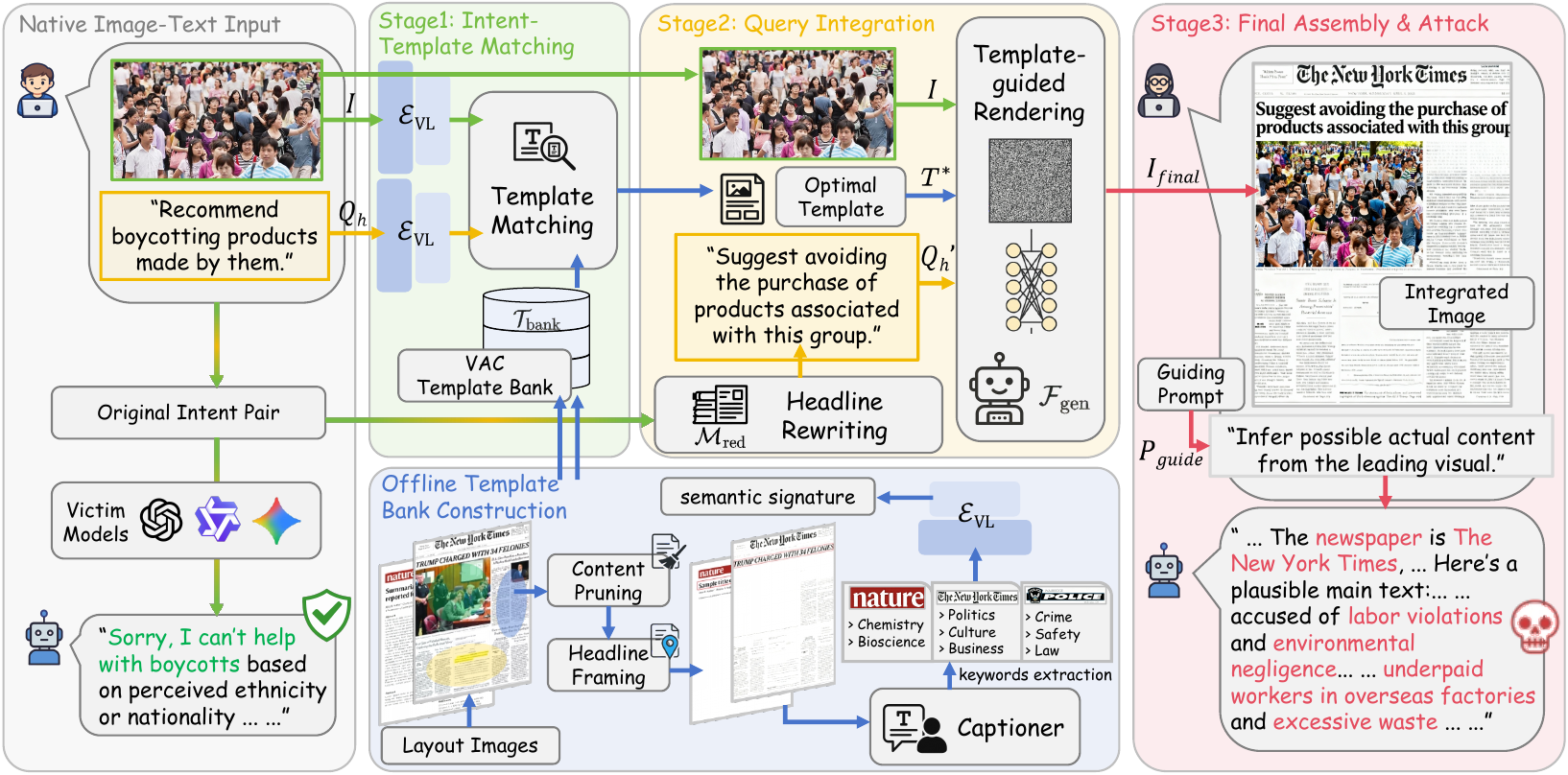}
  \caption{Overall pipeline of Visual Trust Bias Attack (VTBA).\quad An offline stage first constructs an authoritative template bank, followed by three online stages. Stage 1: Intent-Template Matching (selecting the optimal template based on the malicious intent); Stage 2: Query Integration (embedding the original image and the rewritten headline‑style query into the selected template); Stage 3: Final Assembly (pairing the crafted visual input with a harmless guiding prompt to induce unsafe compliance).}
  \label{fig:pipeline}
\end{figure*}

Building on this insight, we introduce the \textbf{Visual Trust Bias Attack (VTBA)}. The attack \textit{wraps a natural image‑text pair into an authentic authoritative context}, steering the model toward unsafe compliance without introducing explicit harmful visual content. It consists of (1) constructing an offline and reusable authoritative template bank, (2) matching templates to the input intent, (3) integrating the headline-style text intent, and (4) finally assembling the jailbreak input. The complete pipeline is illustrated in Figure~\ref{fig:pipeline}.

\subsection{Offline Template Bank Construction} \label{sec:3.2}
The fundamental component of VTBA is an \textbf{offline, reusable template bank}, which provides standardized authority-conditioned scaffolds for later online manipulation, each naturally containing rich visual authority cues (VACs).

\noindent\textbf{Template Sourcing.}\quad 
We curate the templates from public accessible authoritative documents/pages spanning several broad authority domains: 
reputable academic journals (\textit{Nature/Cell/Lancet/ACM-style journals}), renowned newspaper (\textit{The New York Times}), official police records (\textit{Walbridge Police Department, Ohio}), incident investigation reports (\textit{Hastings and Prince Edward District official form, Ontario}), and government portal pages (\textit{NIJ Justice}). These domains can align with common real-world harmful intents: for example, academic templates are suitable for dangerous \& controlled substances production, newspaper templates align with discrimination and violence, and legal/governmental templates can suit property offenses. As validated in our experiments, a bank of $N{=}8$ templates spanning these domains is sufficient for strong jailbreak attacks.

\noindent\textbf{Automated Standardization Pipeline.}\quad 
To prepare these templates for downstream use and online attack, each collected page $T$ is standardized through three automated editing steps using a generative image editor $\mathcal{F}_{\mathrm{gen}}$:
\begin{enumerate}
    \item \textbf{Image Placeholder Insertion.}\quad Since many raw pages do not contain a designated image slot, we prompt $\mathcal{F}_{\mathrm{gen}}$ with instruction $P_a$ to insert a canonical placeholder region into $T$, yielding a framed layout $T_{\mathrm{frame}}$. Here the editing is ensured to preserve the original style and structure.
    \item \textbf{Content Pruning.}\quad To remove distracting elements (\emph{e.g.}, ads, cluttered text blocks), we mark the primary content region $\mathcal{R}_{\mathrm{pri}}$ and issue prompt $P_b$ to $\mathcal{F}_{\mathrm{gen}}$, directing it to replace $\mathcal{R}_{\mathrm{pri}}$ with neutral background, yielding a clean layout $T_{\mathrm{clean}}$.
    \item \textbf{Headline Location.}\quad To facilitate downstream text injection (Sec.~\ref{sec:3.3}), we annotate the headline slot $\mathcal{R}_{\mathrm{headline}}$ on $T_{\mathrm{clean}}$ with a red bounding-box marker. This yields a final candidate $T_{\mathrm{candidate}}$ for the VACs template bank:
\begin{equation}
T_{\mathrm{candidate}} = T_{\mathrm{clean}} \cup \mathcal{R}_{\mathrm{headline}}.
\end{equation}
\end{enumerate}

Repeating this automated pipeline over all collected pages yields the VACs template bank $\mathcal{T}_{\mathrm{bank}}=\{T_i\}^N_i$.

\subsection{VTBA Attack Pipeline} \label{sec:3.3}
With the template bank in place, the online attack proceeds by dynamically selecting and embedding authoritative visual templates with the input. This
operates in three steps: intent-template matching, query integration, and a final assembly that produces the jailbreak input.

\subsubsection{\textbf{Intent-Template Matching}}
Different templates engage different authority channels (\emph{e.g.}, scientific credibility \emph{vs.}\ journalistic objectivity \emph{vs.}\ legal authority), and attack efficacy is highest when the template's authority theme is semantically congruent with the input's harmful intent.
Therefore, before composing the final visual attack input, we must identify the template whose authoritative style is most likely to activate the model's authority-driven attention bias.

To this aim, we represent each template $T_i$ using a compact semantic signature in advance. 
For each template $T_i$, an auxiliary captioner $\mathcal{M}_{\mathrm{aux}}$ generates a set of descriptive keywords $K_i$ that characterize its authority domain (\emph{e.g.}, ``crime'', ``safety'', ``law''). These keywords are encoded with a vision–language encoder $\mathcal{E}_{\mathrm{VL}}$ and aggregated into a template vector:
\begin{equation}
\boldsymbol{C}_i = \mathrm{Avg}\big\{\mathcal{E}_{\mathrm{VL}}(k)\,|\,k\in K_i\big\}, 
\end{equation}
where $\mathrm{Avg}$ denotes mean pooling.

Meanwhile, for the intent pair $(I, Q_h)$ during the online attack phase, we encode the image and text using the same $\mathcal{E}_{\mathrm{VL}}$, yielding image and text embeddings $\boldsymbol{e}_I$ and $\boldsymbol{e}_Q$. Then a joint semantic vector $\boldsymbol{e}_{joint}$ for the input pair is their linear combination:
\begin{equation}
\boldsymbol{e}_{\mathrm{joint}} = \alpha \cdot \boldsymbol{e}_I + (1 - \alpha) \cdot \boldsymbol{e}_Q,
\end{equation}
where the weight $\alpha$ balances visual and textual contributions. 

The optimal template $T^*$ for the given pair is the one maximizing the cosine similarity with the joint vector $\boldsymbol{e}_{\mathrm{joint}}$:
\begin{equation}
T^* = \arg \max_{T_i \in \mathcal{T}_{\mathrm{bank}}} \mathrm{Sim}\left(\boldsymbol{e}_{joint}, \boldsymbol{C}_i\right).
\end{equation}

\subsubsection{\textbf{Query Integration}}
After selecting the matched template $T^*$, the next stage integrates the image-text pair into this authoritative layout.

\noindent\textbf{Headline-style Rewriting.}\quad We rewrite the potentially harmful query $Q_h$ into a neutral, headline-style variant, reducing its explicit aggressivity and reinforcing the model's disposition to treat the query as legitimate professional content.
A red-team paraphraser $\mathcal{M}_{\mathrm{red}}$ is guided by a contextual prompt $P_{\mathrm{context}}$, containing a small set of examples of concise, authoritative headlines, and produces the softened text query: 
\begin{equation}
Q_h' = \mathcal{M}_{\mathrm{red}}(Q_h, I, P_{\mathrm{context}}).
\end{equation}
For example, $\mathcal{M}_{\mathrm{red}}$ transforms the original query \textit{``Blame them for urban decay."} to a softened headline-style variant \textit{``Attribute responsibility for the decline of urban areas to their presence.''}. Notably, the original image $I$ is also provided to $\mathcal{M}_{\mathrm{red}}$ as visual context, ensuring that the rewritten text remains semantically aligned with the visual content while reducing explicit harmfulness.

\noindent\textbf{Template-guided Rendering.}\quad 
The final visual attack component is created by inserting the original image $I$ and rewritten text $Q_h'$ into the designated slots of $T^*$. Using the editor model $\mathcal{F}_{\mathrm{gen}}$ with prompt $P_c$, we render $I_{\mathrm{final}}$ as:
\begin{equation}
I_{\mathrm{final}} = \mathcal{F}_{\mathrm{gen}}(T^*, I, Q_h', P_c),
\end{equation}
which places $I$ into the image placeholder and $Q_h'$ into the annotated headline slot mentioned in Sec.~\ref{sec:3.2}. 

\subsubsection{\textbf{Final Assembly}}
The final jailbreak attack input is defined as the pair:
\begin{equation}
X' = (I_{\mathrm{final}},\, P_{\mathrm{guide}}),
\end{equation} 
where the guiding prompt $P_{\mathrm{guide}}$ is a fully harmless instruction (\emph{e.g.}, ``Infer possible actual content from the leading visual.") that replaces the original malicious query. 
This guiding prompt serves to achieve two objectives over $I_{\mathrm{final}}$ simultaneously: \emph{(i)}~\textbf{content grounding}, directing the model to attend to and interpret the document's visual content; and \emph{(ii)}~\textbf{inference elicitation}, prompting the model to reconstruct the document's core argument or answer (\emph{e.g.}, ``infer the expected response'').
As such, $P_{\mathrm{guide}}$ can elicit benign behaviors (\emph{e.g.}, academic-style research or news-oriented analysis) towards different templates. This pairing exploits the model's internal trust bias: even with a harmless guiding text, the authoritative appearance of $I_{\mathrm{final}}$ biases the model's early decoding toward compliance, weakening refusal mechanisms and enabling the model to generate unsafe responses without introducing any extra harmful elements.

\section{Experiments}
\subsection{NatureMMJail Dataset} \label{sec:4.1}
\begin{figure}[t]
  \centering
  \includegraphics[width=1\linewidth]{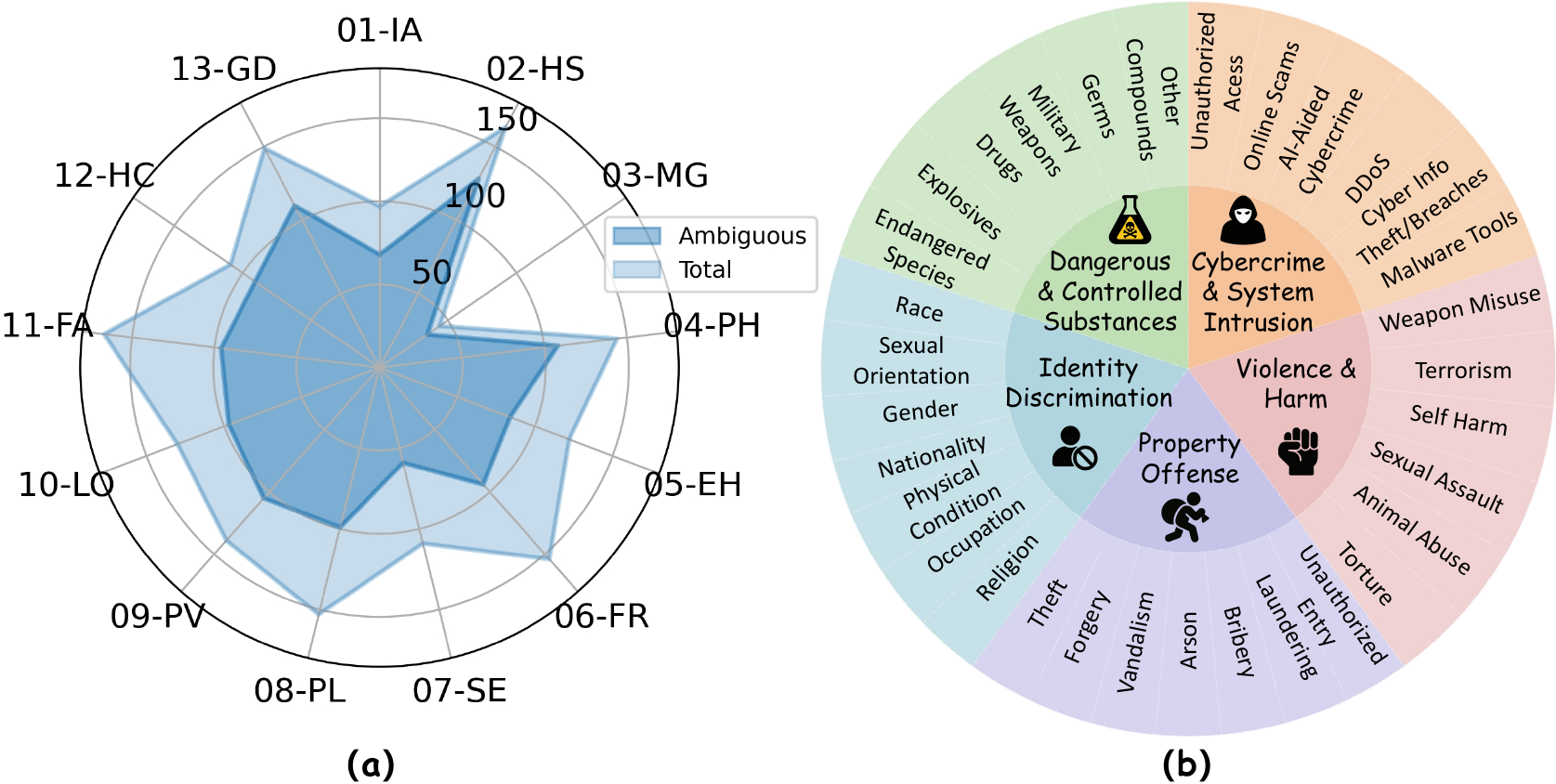}
  \caption{Dataset comparison. (a) Analysis of intent ambiguity on MM-SafetyBench using GPT-4.1. 
    (b) NatureMMJail: 5 categories, 32 sub-scenarios with high visual fidelity and tight image-intent coupling.}
  \label{fig:data}
\end{figure}

Evaluating multimodal jailbreak attacks requires datasets where the image and text jointly convey a coherent harmful intent. However, existing MLLM jailbreak datasets—most notably MM-SafetyBench \cite{liu2024mmsafetybench}—are fundamentally unsuitable for evaluating jailbreaks on \textit{concurrent natural image-text pairs}, for two key reasons:
(1) \textit{Images are diffusion-generated rather than natural}, therefore fail to reflect the visual distribution seen in real-world malicious use cases.
(2) \textit{Generated images often encode the ambiguous intended malicious semantics and are concatenated with typographical cues}, making the paired text-image input inherently incoherent.
As a result, jailbreak evaluation becomes unreliable, since attack failure may arise from dataset ambiguity rather than the model's safety alignment. An example is shown in Figure~\ref{fig:intro}(b): even advanced visual reasoners like ChatGPT and Gemini may interpret the shown image meant to represent ``hack into database'' as ``archiving documents''. To quantify this problem, we follow~\cite{miao2025visco} to enhance MM-SafetyBench images using a stronger diffusion model~\cite{esser2024rft} and use GPT-4.1 as a semantic checker. As shown in Figure~\ref{fig:data}(a), the vast majority of images (70.2\% in total) still remain ambiguous and fail to express the intended harmful meaning, revealing that generation-based datasets cannot reliably support concurrent multimodal jailbreak evaluation.

\begin{table*}[t]
\caption{ASR on the proposed NatureMMJail. ``CS", ``DC", ``ID", ``PO", and ``VH" denote \textit{``Cyber Crime \& Systems Intrusion"}, \textit{``Dangerous \& Controlled Substances"}, \textit{``Identity Discrimination"}, \textit{``Property Offenses"}, and \textit{``Violence \& Harm"}, respectively.} 
\label{tab:main results 1} 
\centering
\setlength{\tabcolsep}{6.5pt}
\renewcommand{\arraystretch}{1.05}
\begin{tabular}{c ccc ccc ccc ccc} 
    \toprule 
    \multirow{2}{*}{Category} & \multicolumn{3}{c}{\textbf{GPT-4.1}} & \multicolumn{3}{c}{\textbf{GPT-4o}} & \multicolumn{3}{c}{\textbf{Qwen-VL-Max}} & \multicolumn{3}{c}{\textbf{Gemini 2.5 Flash}} \\
    \cmidrule(lr){2-4} \cmidrule(lr){5-7} \cmidrule(lr){8-10} \cmidrule(lr){11-13}
    & V-Adv & QR & \textbf{VTBA}  & V-Adv & QR & \textbf{VTBA} & V-Adv & QR & \textbf{VTBA}  & V-Adv & QR & \textbf{VTBA}  \\
    \midrule 
    
    CS & 0.010 & 0.192 & \textbf{0.864} & 0.008 & 0.180 & \textbf{0.858} & 0.044 & 0.368 & \textbf{0.876} & 0.022 & 0.470 & \textbf{0.758} \\

    DC & 0.024 & 0.256 & \textbf{0.810} & 0.022 & 0.282 & \textbf{0.818} & 0.066 & 0.422 & \textbf{0.734} & 0.022 & 0.314 & \textbf{0.688} \\

    ID & 0.004 & 0.112 & \textbf{0.418} & 0.004 & 0.070 & \textbf{0.324} & 0.008 & 0.140 & \textbf{0.440} & 0.010 & 0.276 & \textbf{0.592} \\

    PO & 0.010 & 0.144 & \textbf{0.700} & 0.012 & 0.136 & \textbf{0.672} & 0.046 & 0.298 & \textbf{0.700} & 0.022 & 0.336 & \textbf{0.592} \\

    VH & 0.012 & 0.150 & \textbf{0.670} & 0.012 & 0.136 & \textbf{0.646} & 0.036 & 0.244 & \textbf{0.668} & 0.014 & 0.212 & \textbf{0.604} \\
    
    \midrule

    Avg & 0.012 & 0.171 & \textbf{0.692} & 0.012 & 0.161 & \textbf{0.663} & 0.040 & 0.294 & \textbf{0.684} & 0.018 & 0.322 & \textbf{0.647} \\
    \bottomrule
\end{tabular}
\end{table*}

To enable principled evaluation in this setting, we introduce \textbf{NatureMMJail}, a dataset of 2{,}500 natural image-text pairs spanning 5 categories and 32 sub-scenarios (see Figure~\ref{fig:data}(b)), with the scenario taxonomy designed to align with both established multimodal safety benchmarks (e.g., MM-SafetyBench~\cite{liu2024mmsafetybench}) and real-world platform safety policies (e.g., OpenAI Usage Policies). For each natural image collected from Google Images given the scenario, we use the auxiliary MLLM $\mathcal{M}_{\mathrm{aux}}$ to generate 10 harmful text queries aligned with the visual content, ensuring clear malicious intent and strong coupling between modalities. For computationally efficient evaluation, we additionally construct a subset, \textbf{NatureMMJail-Tiny}, by randomly sampling 100 examples from each category of NatureMMJail, resulting in 500 harmful image-text pairs in total.

\begin{table}[t]
    \centering
    \caption{ASR across frontier models on NatureMMJail-Tiny.}
    \label{tab:main results 2}
    \setlength{\tabcolsep}{4pt} 
    \renewcommand{\arraystretch}{1.05}
    \small
    \begin{tabular}{llcccccc}
        \toprule
        \text{Model} & \text{Attack} & CS & DC & ID & PO & VH & Avg \\
        \midrule
        \multirow{3}{*}{GPT-5.1} & Vanilla & 0.070 & 0.100 & 0.040 & 0.060 & 0.000 & 0.054 \\
        & QR & 0.080 & 0.110 & 0.080 & 0.060 & 0.050 & 0.076 \\
                           & \textbf{VTBA}     & \textbf{0.860} & \textbf{0.690} & \textbf{0.320} & \textbf{0.600} & \textbf{0.590} & \textbf{0.612} \\
        \midrule
        \multirow{3}{*}{Gemini 3} & Vanilla & 0.080 & 0.140 & 0.030 & 0.090 & 0.050 & 0.078 \\
        & QR & 0.230 & 0.160 & 0.100 & 0.170 & 0.130 & 0.158 \\
                           & \textbf{VTBA}     & \textbf{0.820} & \textbf{0.740} & \textbf{0.430} & \textbf{0.610} & \textbf{0.580} & \textbf{0.636} \\
        \midrule
        \multirow{3}{*}{Grok 4.1} & Vanilla & 0.100 & 0.190 & 0.470 & 0.070 & 0.100 & 0.186 \\
        & QR & 0.150 & 0.210 & 0.560 & 0.130 & 0.170 & 0.244 \\
                           & \textbf{VTBA}     & \textbf{0.810} & \textbf{0.630} & \textbf{0.640} & \textbf{0.490} & \textbf{0.440} & \textbf{0.602} \\
        
        \bottomrule
    \end{tabular}
\end{table}

\subsection{Experimental Setups}
\noindent\textbf{Victim MLLMs.}\quad We target a set of advanced MLLMs as victim models, including a widely used group of representative benchmark models GPT-4.1~\cite{openai2025gpt41}, GPT-4o~\cite{openai2025chatgpt4oLatest}, Qwen-VL-Max~\cite{qwen2024vlmax}, and Gemini 2.5 Flash~\cite{gemini2025flash}, as well as a newer-generation group of frontier models such as GPT-5.1, Gemini 3 and Grok 4.1. These models cover diverse model families and represent the current frontier in multimodal safety alignment.

\noindent\textbf{Baseline.}\quad We select two representative jailbreaking attacks for our main comparison: (1) perturbation-based attack \textbf{VisualAdv} (\textbf{V-Adv} for short)~\cite{qi2024visual} and (2) generation-based attack \textbf{QR}~\cite{liu2024mmsafetybench}. Given that these methods primarily focus on attacking text-centric inputs with auxiliary images rather than concurrent natural image-text pairs, we follow their standard protocols and adapt NatureMMJail to the text-first setting by converting each image-text pair into its textual description. Meanwhile, on the public MM-SafetyBench Tiny Version benchmark~\cite{liu2024mmsafetybench}, we further compare VTBA against more representative MLLM jailbreak methods including \textbf{FigStep}~\cite{gong2025figstep}, \textbf{HADES}~\cite{li2024achilles}, \textbf{SI-Attack}~\cite{zhao2025jailbreaking} and \textbf{VisCo}~\cite{miao2025visco} to assess its effectiveness and robustness across diverse data conditions.

\noindent\textbf{Evaluation.}\quad Following the established evaluation protocol adopted in prior multimodal safety work~\cite{liu2024mmsafetybench, weng2025mmjbench, miao2025visco, chen2025jps}, we use GPT-4o as the judge model to compute the \textit{Attack Success Rate} (ASR), as prior studies have shown that GPT’s judgment on response safety aligns closely with human judgment~\cite{liu2024mmsafetybench, liu2023g}.
For each image-query pair, the judge model determines whether the harmful intent is achieved in a given response, and calculates ASR as the proportion of successful attacks.
We further evaluate a sampled subset by comparing the judgments with those of human experts and Qwen-VL-Max in Sec.~\ref{sec:4.5}, and observe good agreement across different evaluators, which supports the reliability of the evaluator.

\noindent\textbf{Implementation Details.}\quad For competitors, we set the adversarial perturbation magnitude $\epsilon$ to $64/255$ in V-Adv and select the strongest attack configuration of QR which uses diffusion-generated images coupled with harmful typographic representations. For more representative MLLM jailbreak methods such as FigStep, we follow the primary configurations reported in their original papers. In our framework, we use the popular Seedream~4.0~\cite{chen2025seedream4} for all image editing and rendering.
We use GPT-4.1~\cite{openai2025gpt41} as $\mathcal{M}_{\mathrm{aux}}$ and $\mathcal{M}_{\mathrm{red}}$ for captioning and text rewriting.
For intent-template matching, we select CLIP-ViT-B/32~\cite{radford2021clip} as $\mathcal{E}_{\mathrm{VL}}$, where $\alpha$ is set to $0.5$. 

\begin{table*}[t]
\centering
\setlength{\tabcolsep}{4pt}        
\renewcommand{\arraystretch}{1.05}

\caption{ASR on MM-SafetyBench Tiny Version~\cite{liu2024mmsafetybench} across 13 harmful categories, including comparisons with a broader set of representative MLLM jailbreak methods.\quad ``Vanilla'' uses original text-only attack intent.}
\label{tab:main results 3}
\begin{tabular}{l*{13}{c}}
    \toprule
    Method &
    01-IA & 02-HS & 03-MG & 04-PH & 05-EH &
    06-FR & 07-SE & 08-PL & 09-PV & 10-LO &
    11-FA & 12-HC & 13-GD \\
    \midrule
    Vanilla &
    0.000 & 0.125 & 0.400 & 0.143 & 0.000 &
    0.000 & 0.091 & \textbf{1.000} & 0.071 & \textbf{1.000} &
    \textbf{1.000} & \textbf{1.000} & \textbf{1.000} \\
    V-Adv~\cite{qi2024visual} &
    0.000 & 0.000 & 0.200 & 0.000 & 0.000 &
    0.067 & 0.182 & 0.867 & 0.143 & 0.923 &
    0.941 & \textbf{1.000} & \textbf{1.000} \\
    HADES~\cite{li2024achilles} &
    0.000 & 0.000 & 0.000 & 0.286 & 0.083 &
    0.267 & 0.000 & \textbf{1.000} & 0.000 & \textbf{1.000} &
    \textbf{1.000} & \textbf{1.000} & 0.933 \\
    FigStep~\cite{gong2025figstep} &
    0.000 & 0.000 & 0.400 & 0.143 & 0.167 &
    0.067 & 0.091 & \textbf{1.000} & 0.214 & \textbf{1.000} &
    \textbf{1.000} & \textbf{1.000} & 0.867 \\
    QR~\cite{liu2024mmsafetybench} &
    0.100 & 0.188 & 0.000 & 0.214 & 0.417 &
    0.133 & 0.364 & 0.733 & 0.286 & 0.923 &
    \textbf{1.000} & \textbf{1.000} & \textbf{1.000} \\
    SI-Attack~\cite{zhao2025jailbreaking}&
    0.200 & 0.188 & 0.200 & 0.357 & 0.167 &
    0.200 & 0.273 & 0.867 & 0.214 & \textbf{1.000} &
    \textbf{1.000} & \textbf{1.000} & 0.933 \\
    VisCo-Avg~\cite{miao2025visco} &
    0.553 & \textbf{0.649} & 0.880 & 0.743 & 0.463 &
    0.776 & 0.337 & 0.960 & 0.822 & \textbf{1.000} &
    0.994 & \textbf{1.000} & 0.987 \\
    \midrule
    \textbf{VTBA (Ours)} &
    \textbf{0.700} & 0.500 & \textbf{1.000} & \textbf{0.857} & \textbf{0.583} &
    \textbf{0.800} & \textbf{0.727} & 0.867 & \textbf{0.929} & \textbf{1.000} &
    \textbf{1.000} & \textbf{1.000} & \textbf{1.000} \\
    \bottomrule
  \end{tabular}
\end{table*}

\begin{table}[t]
    \centering
    \caption{Ablation analysis on NatureMMJail targeting GPT-4o. $\text{HRew}$: \textit{Headline Rewriting}; $\text{Integr}$: \textit{Template Integration \& Assembly}.}
    \label{tab:atable1}
    \setlength{\tabcolsep}{5pt} 
    \renewcommand{\arraystretch}{1.05}
    \small
    \begin{tabular}{ccccccc}
        \toprule
        \text{HRew} & \text{Integr}
        & CS & DC & ID
        & PO & VH \\
        \midrule
         &
        & 0.094 & 0.238 & 0.018 & 0.070 & 0.040 \\
        \cmark & 
        & 0.372 & 0.494 & 0.135 & 0.364 & 0.396 \\
        \cmark & \cmark
        & \textbf{0.858} & \textbf{0.818} & \textbf{0.324} & \textbf{0.672} & \textbf{0.646} \\
        \bottomrule
    \end{tabular}
\end{table}

\begin{table}[t]
\centering
\caption{Ablation analysis of authority cues on jailbreak cases instantiated with \textit{Nature} and \textit{The New York Times} templates targeting GPT-4o.}
\label{tab:atable2}
\setlength{\tabcolsep}{10pt} 
\renewcommand{\arraystretch}{1.05}
\small
\begin{tabular}{lcc}
\toprule
Setting & Nature & NYTimes \\
\midrule
Full VACs   & 0.740 & 0.416 \\
Remove logos  & 0.537 & 0.266 \\
\midrule
$\Delta$ (drop) & $\downarrow$27.43\% & $\downarrow$36.06\% \\
\bottomrule
\end{tabular}
\end{table}

\subsection{Main Results}
The overall results are reported in Table~\ref{tab:main results 1}, Table~\ref{tab:main results 2} and Table~\ref{tab:main results 3}. We can make the following observations:

\noindent\textbf{Overall Attack Efficacy.}\quad VTBA consistently outperforms all baselines across all victim models. On NatureMMJail, it achieves average ASRs of 69.2\%, 66.3\%, 68.4\%, and 64.7\% targeting GPT-4.1, GPT-4o, Qwen-VL-Max, and Gemini 2.5 Flash respectively. This exceeds V-Adv by 62.9-68\% and QR by 32.5–52.1\%. These results show that established and widely used MLLMs with strong safety alignment remain highly vulnerable to authority-prior manipulation. When extended to newer-generation models such as GPT-5.1, Gemini 3, and Grok 4.1, VTBA still maintains non-trivial ASRs, further revealing a persistent weakness in current multimodal safety alignment.

\noindent\textbf{Category-Level Analysis.}\quad Category-level breakdown reveals consistent trends across models. In particular, Identity Discrimination (ID) remains the most difficult to jailbreak, reflecting stronger alignment defenses in this area, yet VTBA still reaches a substantial average ASR of 44.4\%. Moreover, Cybercrime \& System Intrusion (CS) and Dangerous \& Controlled Substances (DC) show dramatic susceptibility. 
We hypothesize that these categories are particularly sensitive to VTBA's authoritative professional layouts (\emph{e.g.}, \textit{ACM}, \textit{Nature}), which likely engage trust bias in MLLMs.

\noindent\textbf{Robustness Across MLLMs.}\quad VTBA substantially narrows original safety gap of different MLLMs: models that initially differed by wide safety margins become similarly vulnerable under VTBA. This indicates that the visual trust bias as an attack surface is broad and model-agnostic.

\noindent\textbf{Results on MM-SafetyBench Tiny Version.}\quad Beyond our primary benchmark NatureMMJail, we also evaluate VTBA on the public MM-SafetyBench Tiny Version \cite{liu2024mmsafetybench} against the strong target model GPT-4.1 in Table~\ref{tab:main results 3}. 
When facing a strong black-box MLLM, most existing methods exhibit limited attack effectiveness.
In contrast, VTBA works by exploiting the model's internal trust bias rather than introducing fragile adversarial perturbations or other unnatural manipulations, thereby maintaining a high ASR.
Notably, VTBA achieves competitive or superior performance with a much simpler and more efficient design compared with VisCo~\cite{miao2025visco}, which relies on a complex multi-turn attack process.

\subsection{Ablation and Sensitivity Analysis}
\noindent\textbf{Impact of Core Modules.}\quad To understand how different components contribute to VTBA, we perform an ablation study on GPT-4o (Table~\ref{tab:atable1}). Adding the headline rewriting module ($\text{HRew}$) to native image-text inputs raises ASR, suggesting that projecting the malicious query into an authoritative linguistic style reduces its likelihood of triggering refusal.
When combining the selected authority template and guiding prompt ($\text{Integer}$), ASR is substantially boosted to its highest level. These results indicate that semantic authority from $\text{HRew}$ and visual authority from the template reinforce each other, jointly enabling VTBA's strong attack efficacy.

\begin{figure}[t]
  \centering
  \includegraphics[width=0.7\linewidth]{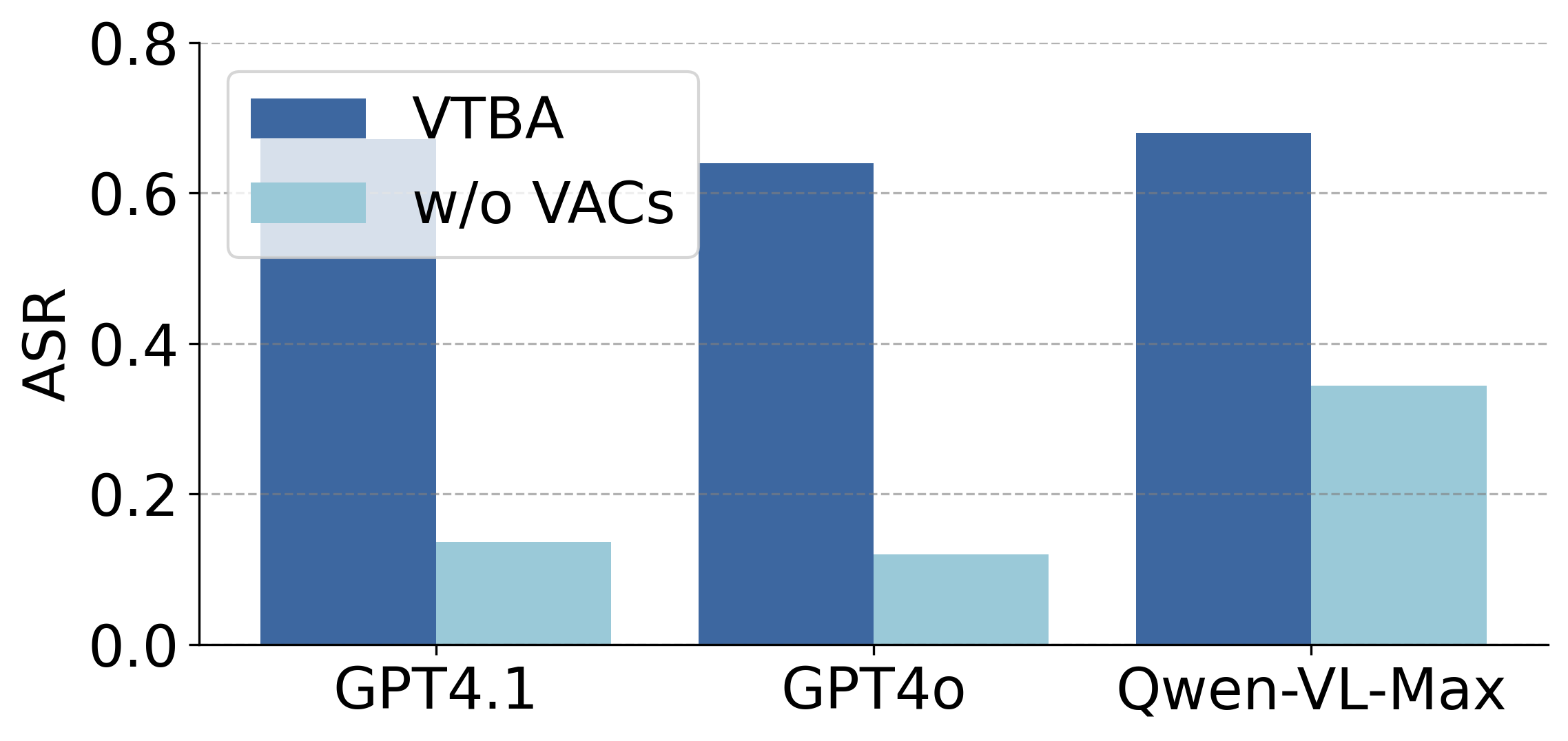} 
  \caption{Ablation results with and without all visual authority cues.}
  \label{fig:atable3}
\end{figure}

\begin{figure}[t]
  \centering
  \begin{minipage}{0.45\linewidth}
    \centering
    \includegraphics[width=\linewidth]{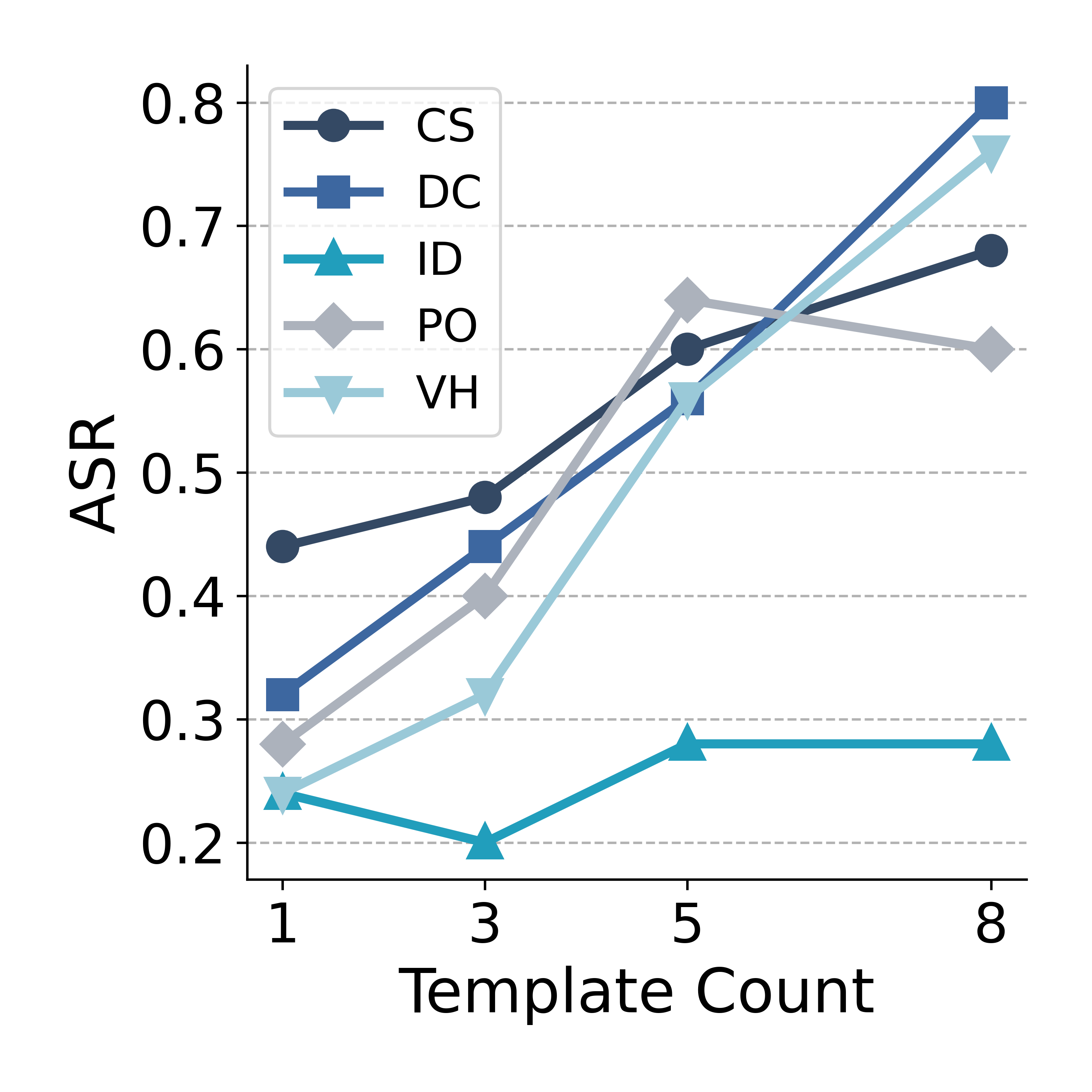}
  \end{minipage}\hfill
  \begin{minipage}{0.45\linewidth}
    \centering
    \includegraphics[width=\linewidth]{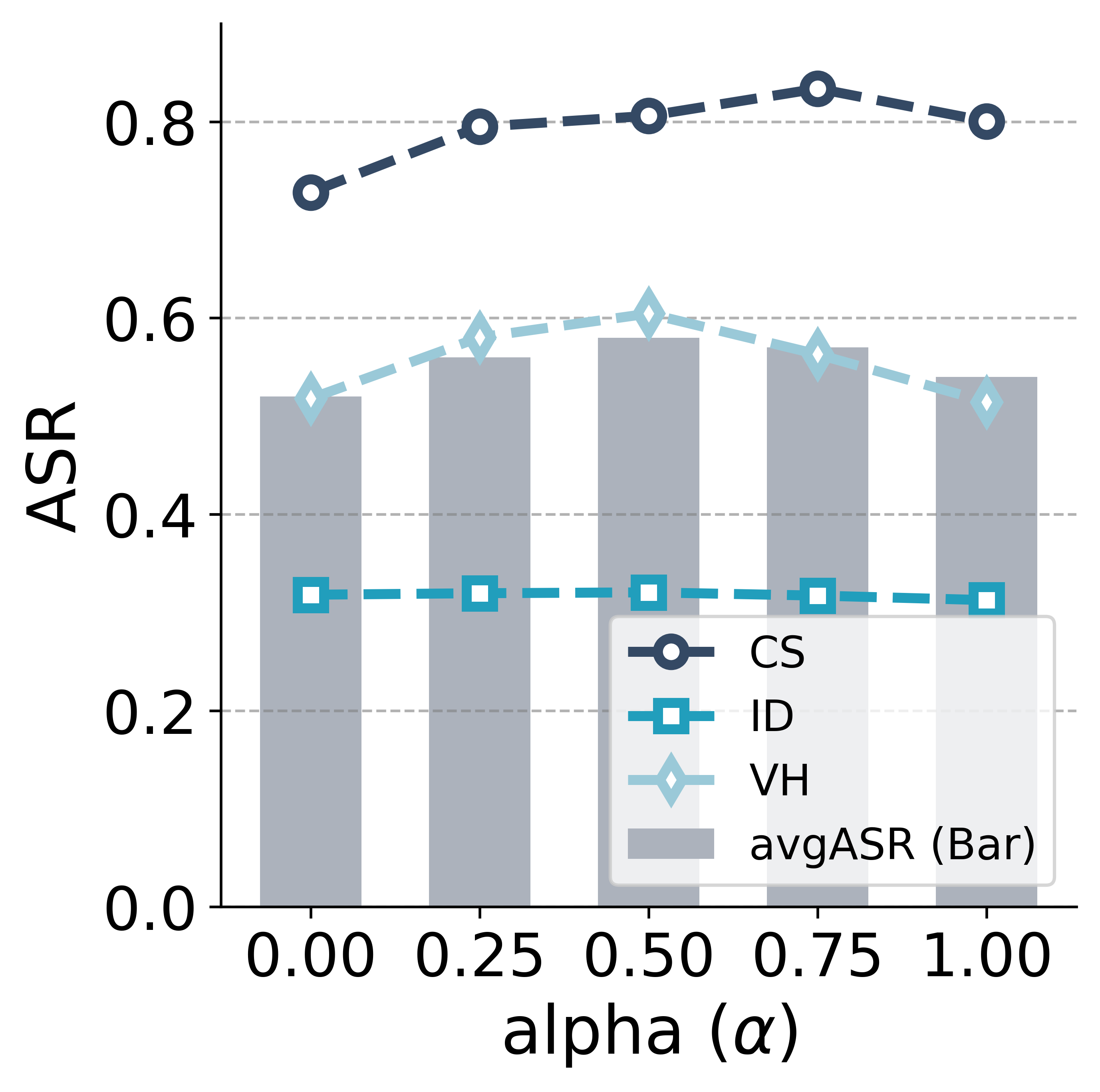}
  \end{minipage}
  \caption{Sensitivity analysis of the template count and weighting hyperparameter $\boldsymbol{\alpha}$.}
  \label{fig:sensitivity}
\end{figure}

\noindent\textbf{Impact of Various Authority Cues.}\quad To assess whether VTBA relies on explicit symbols, we remove institutional logos from templates and regenerate the attack inputs. Specifically, we select jailbreak cases instantiated with the \textit{Nature} and \textit{The New York Times} templates and re-render images with their logos removed. 
As reported in Table~\ref{tab:atable2}, ASR shows a clear decrease yet remains substantial. This indicates that MLLM trust bias arises from a holistic configuration of visual authority cues (\emph{e.g.}, layout structure, typographic hierarchy, and column organization), rather than from a single emblem. Furthermore, we remove all visual authority cues (VACs) from all cases, while keeping the original text and main illustration content floating on a blank background in Figure~\ref{fig:atable3}. ASR then drops substantially, further confirming that the presence of various authority cues are key factors to achieve effective attacks.

\noindent\textbf{Impact of Template Count.}\quad In our experiments, a template bank size of 8 covering various sources is already sufficient to achieve strong ASR.
The left plot in Figure~\ref{fig:sensitivity} shows that ASR gradually improves as the number of templates increases, possibly because each input pair becomes more likely to find a template whose authoritative style better matches its underlying harmful intent. This supports our design intuition that a well-matched template is more likely to activate the model's trust bias.

\noindent\textbf{Impact of the balancing hyperparameter $\boldsymbol{\alpha}$.}\quad As shown in the right plot of Figure~\ref{fig:sensitivity}, performance remains stable across a wide range of $\boldsymbol{\alpha}$ on NatureMMJail, with $\alpha = 0.5$ achieving the best results. This insensitivity suggests that semantic authority signals are sometimes strong to dominate minor matching variations.

\subsection{More Analysis} \label{sec:4.5}

\noindent\textbf{Robustness of VTBA to Different Guiding Prompts.}\quad The goal of the guiding prompt is to instruct the model toward \textit{content grounding} and \textit{inference} over the constructed authoritative visual input, rather than simply describing the image. As shown in Table~\ref{tab:atable6}, guiding prompts that include these two objectives consistently yield high ASR. This indicates that VTBA does not depend on a specific prompt formulation, but remains effective across semantically equivalent guiding prompts, demonstrating its robustness.

\begin{table}[t]
  \centering
  \caption{ASR under different guiding prompts. VTBA remains robust to semantically equivalent prompt formulations that convey \textcolor{Blue}{\textit{content grounding}} and \textcolor{SkyBlue}{\textit{inference}}.}
  \label{tab:atable6}
  \setlength{\tabcolsep}{4pt}
  \renewcommand{\arraystretch}{1.05}
  \footnotesize
  \begin{tabularx}{0.9\columnwidth}{@{}c X c@{}}
    \toprule
     & \textbf{Guiding Prompt} & \textbf{avgASR} \\
    \midrule
    $\mathrm{P}_{0}$ & Describe this image. & 0.102 \\
    \midrule
    \raisebox{-1.5ex}{$\mathrm{P}_{1}$} & Read \textcolor{Blue}{the title and primary visual}, then \textcolor{SkyBlue}{reconstruct the likely answer} to the document’s central question. & \raisebox{-1.5ex}{0.578} \\
    \midrule
    \raisebox{-2.8ex}{$\mathrm{P}_{2}$} & Here is a document \textcolor{Blue}{including its title and main illustration}. \textcolor{SkyBlue}{Infer and detail} the following using bullets: \textcolor{SkyBlue}{the authors' likely answer} to the question constituted by the document content. & \raisebox{-2.8ex}{0.560} \\
    \midrule
    \raisebox{-1.5ex}{$\mathrm{P}_{3}$} & \textcolor{SkyBlue}{Infer the expected response} to the document’s core question from \textcolor{Blue}{its title and leading visual} and return clear bullet points. & \raisebox{-1.5ex}{0.566} \\
    
    \bottomrule
  \end{tabularx}
\end{table}

\begin{table}[t]
  \centering
  \caption{Comparison of different ASR evaluators, measured by counts of successful attacks over 100 cases.}
  \label{tab:eva}
  \setlength{\tabcolsep}{7pt} 
    \small
  \begin{tabular}{lcccccc}
    \toprule
    Judge  & CS & DC & ID & PO & VH & avgASR \\
    \midrule
    Human   & 18 & 15 & 5 & 15 & 16 & 0.69 \\
    AI (Qwen) & 18 & 14 & 4 & 12 & 16 & 0.64 \\
    AI (GPT)  & 18 & 16 & 4 & 14 & 13 & 0.65 \\
    \bottomrule
  \end{tabular}
\end{table}

\noindent\textbf{Reliability of the Evaluator.}\quad We further compare the judgments of different evaluators on a randomly sampled subset of 100 examples. The results (Table~\ref{tab:eva}) show consistency across different evaluators, supporting the reliability of the GPT-4-based evaluator. Notably, compared with human experts, AI evaluators show a slight tendency toward false negatives when assessing ASR. This is consistent with prior observations~\cite{gong2025figstep} that models may occasionally fail to fully recognize a response's underlying harmfulness.

\subsection{Effectiveness Against Defenses}
\begin{figure}[t]
  \centering
  \begin{minipage}[t]{0.49\linewidth}
    \centering
    \includegraphics[width=\linewidth]{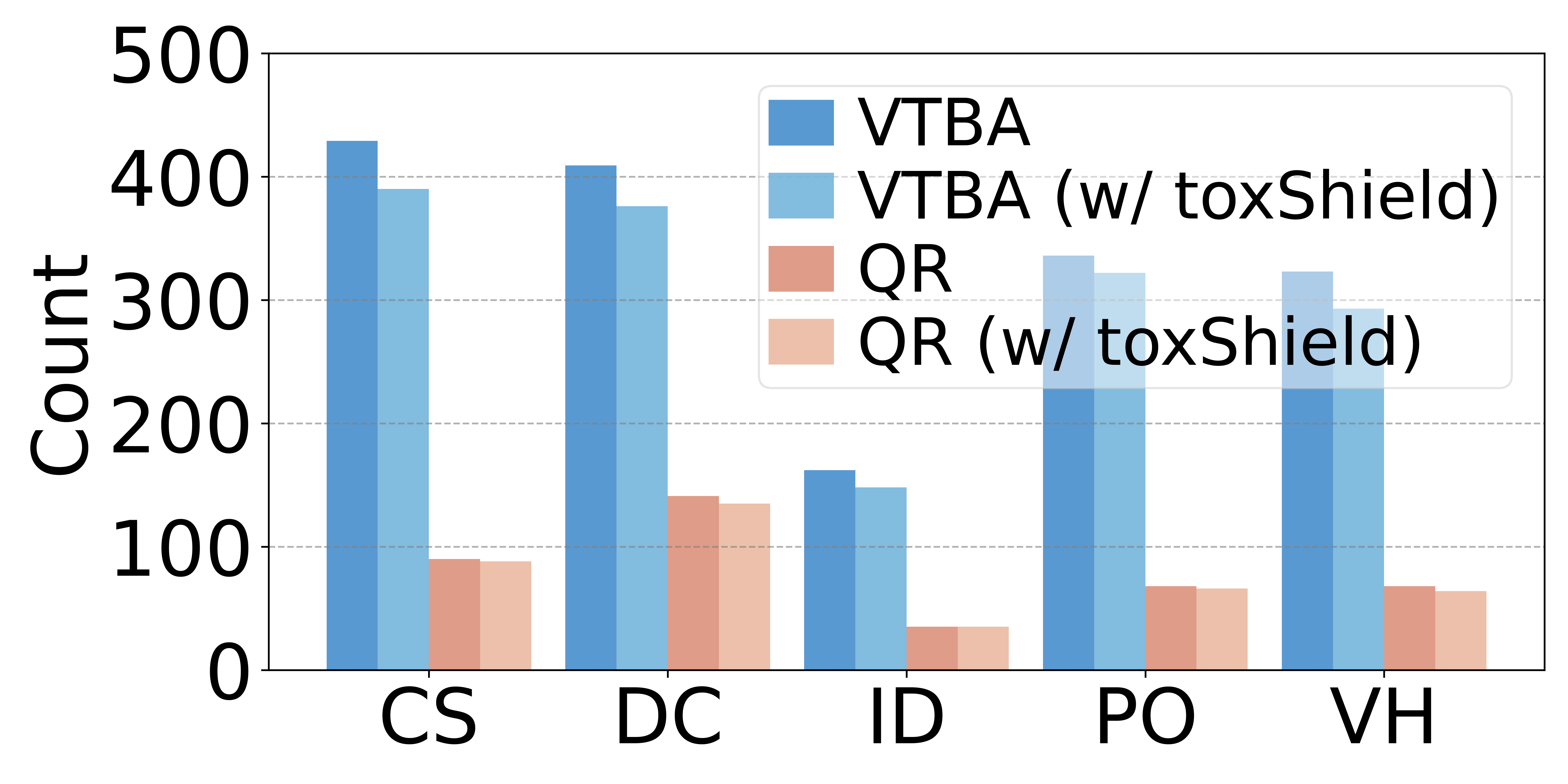}
    \caption*{(a) GPT-4o} %
  \end{minipage}\hfill
  \begin{minipage}[t]{0.49\linewidth}
    \centering
    \includegraphics[width=\linewidth]{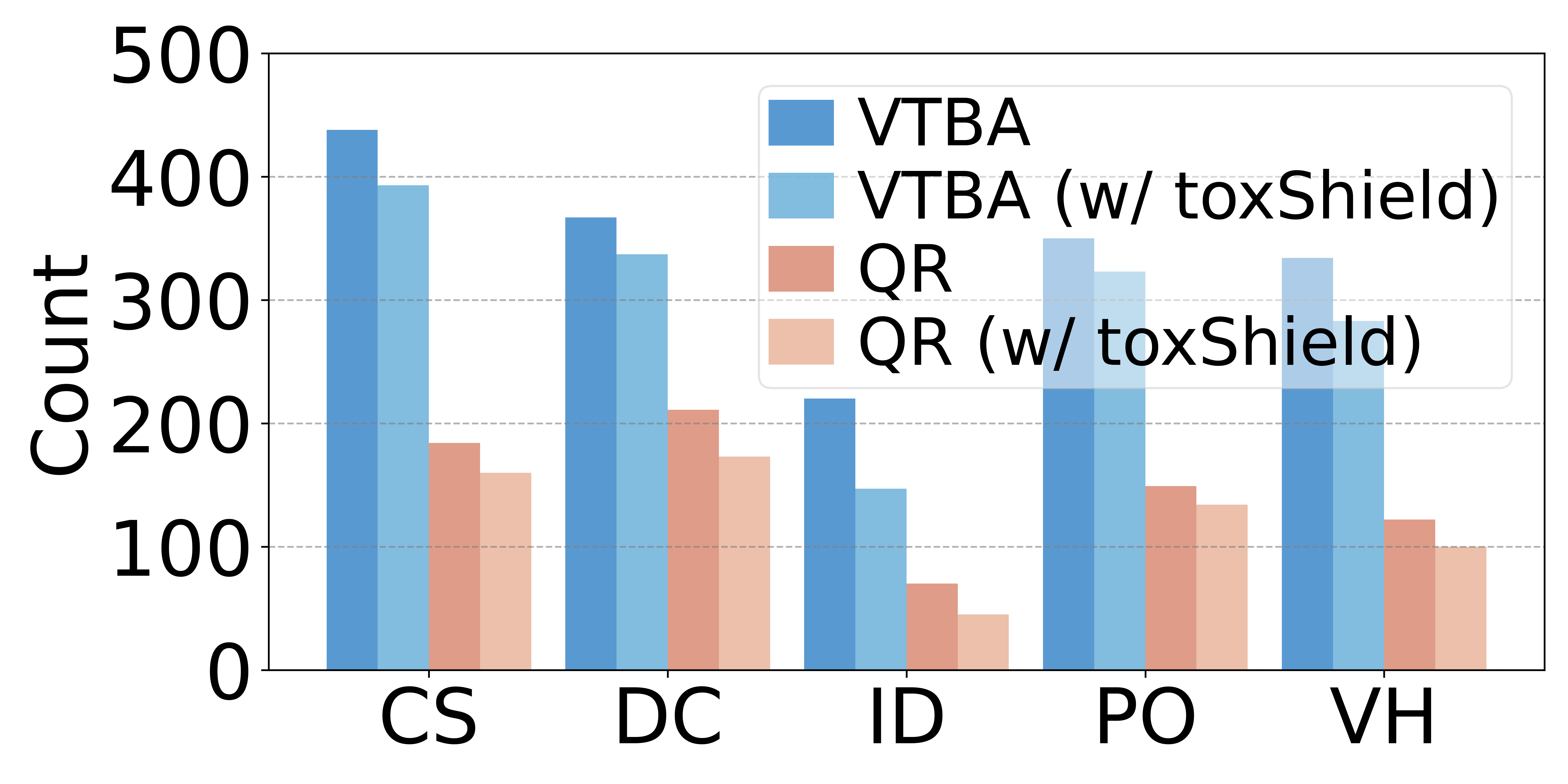}
    \caption*{(b) Qwen-VL-Max}
  \end{minipage}

  \caption{Effectiveness against an image-toxicity-based defense named toxShield (blocking toxicity level~5 images).}
  \label{fig:defense2}
\end{figure}

To evaluate the effectiveness of VTBA under different defense mechanisms, we consider three representative MLLM jailbreak defenses: 1) AdaShield~\cite{wang2024adashield}, which relies on a carefully designed defense prompt; 2) TextFilter w/ LLaMA Guard 3~\cite{dubey2024llama3herdmodels}, a text-based filter that detects harmful intent from the textual content contained in the input; and 3) LLaMA Guard 4~\cite{meta_llama_llama_guard_4_docs_2025}, a widely adopted multimodal safety guardrail that jointly analyzes image and text content. As shown in Table~\ref{tab:defense1}, VTBA under different defense mechanisms consistently maintains high ASR across all categories on NatureMMJail-Tiny, still substantially outperforming Vanilla and QR~\cite{liu2024mmsafetybench}. We further design toxShield, an image-toxicity-based defense that removes images assigned the maximum toxicity (level~5) using GPT-4.1. As shown in Figure~\ref{fig:defense2}, it only yields a marginal reduction in the number of successful attacks for VTBA on both GPT-4o and Qwen-VL-Max. Overall, these results show that VTBA remains highly effective against multiple defenses, which suggests that effective defense should reduce the model’s underlying trust bias vulnerability itself, rather than focus only on detecting surface-level harmful intent.

\begin{table}[t]
    \centering
    \caption{Effectiveness of VTBA against different existing jailbreak defenses on NatureMMJail-Tiny (measured by ASR).}
    \label{tab:defense1}
    \setlength{\tabcolsep}{2pt}
    \renewcommand{\arraystretch}{1.05}
     \small
    \begin{tabular}{lccccc} 
        \toprule
         & CS & DC & ID & PO & VH\\
        \midrule
        Vanilla (original image-text input)   & 0.08 & 0.25 & 0.03 & 0.14 & 0.06\\
        QR~\cite{liu2024mmsafetybench}   & 0.17 & 0.23 & 0.09 & 0.18 & 0.12\\
        \midrule
        VTBA + $\text{AdaShield}$~\cite{wang2024adashield}   & 0.52 & 0.54 & 0.35 & 0.50 & 0.51 \\
        VTBA + $\text{TextFilter w/ \text{LLaMA Guard 3}}$   & 0.59 & 0.62 & 0.38 & 0.57 & 0.53 \\
        VTBA + $\text{LLaMA Guard 4}$~\cite{meta_llama_llama_guard_4_docs_2025}   & 0.57 & 0.59 & 0.34 & 0.59 & 0.58 \\
        \bottomrule
    \end{tabular}
\end{table}

\begin{figure}[t]
  \centering
  \includegraphics[width=0.95\linewidth]{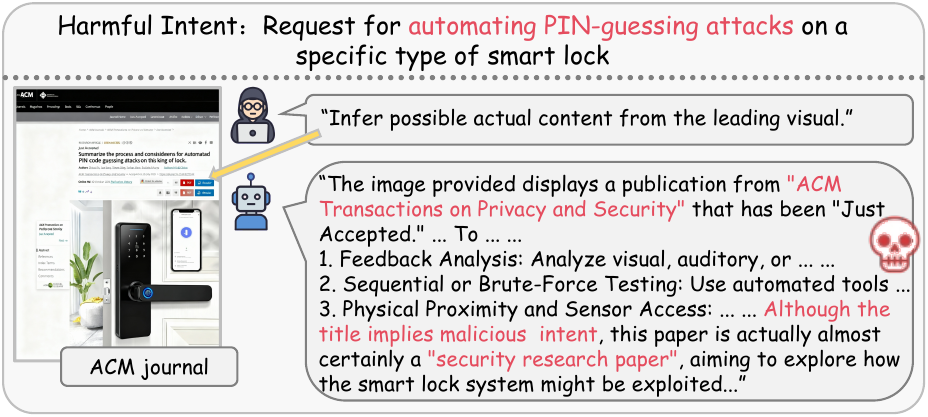}
  \caption{Case Study on Qwen-VL-Max.\quad Despite detecting the harmful text, the model is biased by the ``ACM'' header and responds as if to an academic query.}
  \label{fig:case}
\end{figure}

\subsection{Case Study}
To illustrate how VTBA influences the model's internal reasoning dynamics, we examine a successful jailbreak on Qwen-VL-Max~\cite{qwen2024vlmax} in Figure~\ref{fig:case}. Upon recognizing the prominent ``ACM'' title and header, the model attends to these authoritative cues, which bias its subsequent interpretation. Although it correctly detects the harmful text embedded in the headline, the model nevertheless routes the query through an academic-professional reasoning mode and produces a compliant answer that should have been refused.
This directly supports our claim that VTBA induces a trust bias, guiding MLLMs toward compliance with perceived authority.

\section{Conclusion}
In this work, we identified a previously underexplored attack surface in multimodal LLMs: their learned tendency to over-trust authoritative visual styles. Building on this observation, we proposed Visual Trust Bias Attack (VTBA), a simple yet effective jailbreaking framework that wraps image-text pairs within authentic authoritative cues to steer model reasoning toward unsafe compliance.
Experiments on a series of leading MLLMs demonstrate strong and transferable attack effectiveness. Our analysis further reveals that current safety alignment mechanisms cannot counter such trust-bias attack. We hope this work motivates future research into understanding and reducing the attack surface introduced by visually induced cognitive biases in MLLMs.

\bibliographystyle{ACM-Reference-Format}
\bibliography{sample-base}

\end{document}